\documentstyle[aps,prb,tighten]{revtex}

\begin{document}

\twocolumn[

\hsize\textwidth\columnwidth\hsize
\csname@twocolumnfalse\endcsname

\draft
\title{``Symmetry properties of magnetization in the Hubbard model at
finite temperature'': Comment on `Reply to Comment'}
\author{A.~Avella$^\dag$, F.~Mancini$^\dag$ and D.~Villani$^{\dag,\ddag}$}
\address{$^\dag$Universit\`a degli Studi di Salerno --- Unit\`a INFM di Salerno\\
Dipartimento di Scienze Fisiche ``E.~R.~Caianiello'', 84081 Baronissi,
Salerno, Italy\\$^\ddag$Serin Physics Laboratory, Rutgers University,
Piscataway, New Jersey 08854, USA}
\date{\today}
\maketitle

\begin{abstract}
We comment on the Reply of G.~Su and M.~Suzuki
{[\emph{cond-mat}/9808011]} to our Comment {[\emph{cond-mat}/9807402]}
about the papers of G.~Su {[Phys.~Rev.~B
\textbf{54}, 8291 (1996)]} and G.~Su and M.~Suzuki {[Phys.~Rev.~B
\textbf{57}, 13367 (1998)]}. The algebraic mistake originating their
incorrect solutions is evidenced in detail.
\end{abstract}
\pacs{}]

In our Comment\cite{one} to the papers of G.~Su\cite{two} and G.~Su and
M.~Suzuki\cite{three} we questioned the solutions for both spin and
pseudo-spin correlation functions. The Reply\cite{four} has been
articulated in three points which turn out to be all wrong. Also, the
Authors enrich their script with adjectives and tones which are alien to
a consolidated Physics tradition of serious scientific debates. Clearly,
all these points call for a second Comment.
\begin{enumerate}
\item Our main argument regarding the zero temperature case was that
the formalism of the equation of motion is valid for any temperature,
including also zero temperature. In this sense, Eqs.~($5$) and ($7$) of
Ref.~\onlinecite{one} remain true also at zero temperature, where one is
just concerned with expectation values on the ground state, whose more
or less exotic nature cannot affect anyway the framework of
calculations.
\item The Authors promote as a possible expression for the ferromagnetic
magnetization per site $m(h,T)$ in the Hubbard model at finite
temperature what it follows
\begin{equation}
m(h,T)=\frac n2\tanh(\beta h)
\label{eq1}
\end{equation}

To put in stronger evidence this result, they assess that it also agrees
with the two limiting cases of noninteracting [i.e., $U=0$] and
half-filling [i.e., $n=1$] ones. In contrast, the formula (\ref{eq1}) is
certainly wrong for the non-interacting case whereas it holds at
half-filling only in the limit\cite{five} $U\rightarrow\infty$. However,
in the Reply they continue to show a scarce familiarity with
Ref.~\onlinecite{five} and outline a false demonstration for the
noninteracting case. In combining their Eqs.~($5$) and ($6$) there is a
repeated and tautological use of Eq.~($3$) which would produce the
trivial identity
\begin{equation}
m(h,T)=m(h,T)
\label{eq2}
\end{equation}
in absence of a striking algebraic error. In fact, the factor $2$ that
multiplies the second term of the right-hand side of Eq.~($7$) is
actually absent by carrying out correctly analytical calculations.
\item In reply to the comments about their results for the pseudo-spin
correlation function, the Authors stress that our Fig.~$2$ is invalid
being the filling also dependent on $T$, $U$ and $\mu$. Apart the
questionable adjective (e.g., \emph{ridiculous}) in reference to our
Fig.~$2$, it results really difficult to understand their remark, being
the filling thermodynamic independent variable as the temperature and
the on-site Coulomb repulsion.
\end{enumerate}

In conclusion, the whole argument of G.~Su and M.~Suzuki is simply
wrong. Therefore, we reconfirm our results\cite{one} and point out again
that their solutions for both spin and pseudo-spin correlation functions
are incorrect. Besides, where in the text mistakes can be ruled out, it
raises on a verbal attitude that is totally extraneous to the Physics
community. We think that a moderate and not rude presentation is a
necessary precondition for dissemination of scientific results. As a
matter of fact, adjectives as \emph{ridiculous} more than simple-minded
and wrong applications of standard formulas, seem to be not acceptable
in the context of a scientific debate.

\end{document}